\documentclass[conference]{IEEEtran}
\IEEEoverridecommandlockouts
\usepackage{algorithmic}
\usepackage{graphicx}
\usepackage{multirow}
\usepackage{textcomp}
\usepackage{array}
\usepackage{tabularx}
\usepackage{xcolor}
\usepackage{hyperref}
\usepackage{amsfonts}
\usepackage{subcaption}
\usepackage{scalerel}
\usepackage{siunitx}
\def\BibTeX{{\rm B\kern-.05em{\sc i\kern-.025em b}\kern-.08em
    T\kern-.1667em\lower.7ex\hbox{E}\kern-.125emX}}

\usepackage[backend=bibtex,giveninits=true,hyperref=true,maxnames=1,minnames=1,natbib=true,style=savetrees,sorting=none]{biblatex}
\usepackage[moderate,tracking=normal,paragraphs=tight,mathspacing=tight,charwidths=tight]{savetrees}
\usepackage[compact]{titlesec}
\usepackage{cuted} 
\begin{document}

\title{Group Equivariant Convolutional Networks for Pathloss Estimation\\
%{\footnotesize \textsuperscript{*}Note: Sub-titles are not captured in Xplore and
%should not be used}
%\thanks{Identify applicable funding agency here. If none, delete this.}
}

\author{\IEEEauthorblockN{Ziyue Yang\IEEEauthorrefmark{1}, Feng Liu\IEEEauthorrefmark{2}, Yifei Jin\IEEEauthorrefmark{3}, and Konstantinos Vandikas\IEEEauthorrefmark{3}}
\IEEEauthorblockA{\IEEEauthorrefmark{1} KTH Royal Institute of Technology,  Stockholm, Sweden, ziyuey@kth.se}
\IEEEauthorblockA{\IEEEauthorrefmark{2} Ericsson, Beijing, China, feng.c.liu@ericsson.com}
\IEEEauthorblockA{\IEEEauthorrefmark{3} Ericsson Research, Stockholm, Sweden, firstname.lastname@ericsson.com}
}

\maketitle

\begin{abstract}
This paper presents RadioGUNet, a UNet-based deep learning framework for pathloss estimation in wireless communication. Unlike other frameworks, it leverages group equivariant convolutional networks, which are known to increase the expressive capacity of a neural network by allowing the model to generalize to further classes of symmetries, such as rotations and reflections, without the need for data augmentation or data pre-processing. The results of this work are twofold. First, we show that typical UNet-based convolutional models can be easily extended to support group equivariant convolution (\textsc{g-conv}). Secondly, we show that the task of pathloss estimation benefits from such an extension, as the proposed extended model outperforms typical UNet-based models by up to 0.41~\unit{\decibel} for a similar number of parameters~\footnote{In this paper, the term  \textit{parameter} refers to the trainable weights of the neural network (e.g., the convolutional filters). The number of parameters reflects the model’s capacity and its computational cost.} in the RadioMapSeer dataset. The code is publicly available on the GitHub page: https://github.com/EricssonResearch/radiogunet
\end{abstract}

\begin{IEEEkeywords}
CNN, UNet, equivariant neural networks, deep learning, propagation map
\end{IEEEkeywords}

\section{Introduction}
6G networks~\cite{dang2020should} constitute the next generation of mobile networks succeeding 5G. As with every new generation, 6G is expected to outperform its predecessors not only in terms of latency and throughput but also in energy efficiency, allowing enhanced user experience in applications such as AR/XR by offering differentiated connectivity. 6G is to achieve that by leveraging higher frequencies in millimeter-wave and terahertz ranges that come at the expense of denser network deployments.

In the design and implementation of 6G Networks, pathloss estimation becomes an important factor. Pathloss refers to the reduction in signal strength (power) as radio waves propagate from a transmitter to a receiver. It is influenced by factors such as distance, frequency, and environmental conditions. Pathloss estimation is critical for 6G networks, as it is directly related to network performance, coverage, and user experience. 

Pathloss estimation is used in tasks such as network planning and coverage design to determine the optimal placement of base stations, as well as the types of cells that can be smaller in dense areas, such as microcells or femtocells. Moreover, it is used to determine the link budget and signal quality, therefore determining transmission power to ensure reliable connections. It is used to manage interference between adjacent cells that may occur if there are overlaps in the placement of the cells. Particularly for 5G advanced and 6G networks where higher frequencies are used (such as 28 or 39 GHz), higher pathloss is experienced. Therefore, the design of such mobile networks significantly benefits from pathloss estimation to determine the placement of cells.

Pathloss is typically expressed in decibels (\unit{\decibel}) and can be estimated using different types of models taxonomized as theoretical/foundational models, basic models, terrain models, supplementary models, stochastic fading models, many-ray models, and active measurement models~\cite{6165627}. Moreover, different types of neural networks have been explored to estimate pathloss~\cite{ahmadi2025unet, levie2021radiounet, petrosyanu}. A common element of these approaches is that they employ the UNet type of neural architecture~\cite{ronneberger2015u}, which is well known in computer vision and applied to tasks such as image segmentation~\cite{minaee2021image}. 
The primary challenge when using UNet, a convolutional type of neural network, is to create an estimator whose performance does not degrade when asked to estimate data not included in the training dataset. To train such models effectively, a comprehensive dataset is typically required, one that encompasses a wide range of scenarios. In computer vision, this often involves including multiple transformed versions of the same image, such as various rotations and reflections. Analogously, for pathloss estimation, the dataset must capture the diverse propagation phenomena that occur in real environments, including reflections, diffractions, and scattering caused by obstacles. Without sufficient coverage of these effects in the training data, the model's ability to generalize to new environments is limited. In practice, this can be achieved using a high volume of synthetic/simulated data, but this may increase the training time of the model. Alternatively, data augmentation techniques may be employed to achieve the same goal. However, these would also increase the size of the input dataset and, consequently, the training time.

To overcome this challenge, group equivariant neural networks~\cite{cohen2016group} have been proposed. Instead of generating new data to capture different versions of the same input, the kernel of the convolution process is transformed, thus resulting in transformed feature maps (the output of the convolution process), which are identical to those that would have been obtained if the model were trained using various transformations of the input data. As such, group equivariant convolution (\textsc{g-conv}) allows for obtaining models that generalize without the need for additional data~\cite{bekkers2018roto}.

The main contribution of this work is the study and application of \textsc{g-conv} to the problem of pathloss estimation in a UNet-based neural architecture. We demonstrate that extending the UNet architecture with group equivariance improves pathloss estimation accuracy.

The paper is structured as follows: Section~\ref{sec:background} presents relevant work and corresponding challenges in the field of pathloss estimation. Section~\ref{sec:method} introduces the proposed RadioGUNet model and group equivariant neural networks. Section~\ref{sec:dataset} describes the datasets used in our experiments to train and validate the proposed model, while Section~\ref{sec:result} presents and discusses the evaluation results. Finally, in Section~\ref{sec:conclusions}, we conclude this work and present possible enhancements.

\section{Background}
\label{sec:background}
% Pathloss - > Radio Map 
Besides wireless channel modeling, the recent development of pathloss estimation has shifted its focus from \textit{link-level} to \textit{cell-level} scenarios. 
\citet{zhu2024physics} have discussed the diversity of wireless signal distribution under rich environment semantics. They pointed out that shadowing, small-scale fading, and diffraction, which are intricately linked to the complexity of environmental semantics, contribute to the difficulty of pathloss estimation at scale. 
One mainstream method to handle such diversity is employing a prior with environment information (e.g., Channel geometry from wireless ray-tracing~\cite{Jin2505:SANDWICH}, radio coverage scope~\cite{Jin2303:Learning}, or radio angle of arrival \& departure~\cite{zhao2023nerf2}) to assist pathloss estimation. However, due to the presence of extra components (e.g., wireless ray tracer~\cite{hoydis2023sionna}), information (e.g., inter-cell relations) or signaling, such prior knowledge is not always available for macro-scale measurements. The most similar work is \textsc{NeRF$^2$}~\cite{zhao2023nerf2}, which models a spatial prior. However, their supervision is based on a compact integration of the indoor scene, which is intractable for outdoor scenarios. 

% Radio Map -> CNN method -> PINN method
In addition to obtaining assistance from different forms of radio environmental priors, one of the most conventional approaches for pathloss estimation at scale is to adopt computer vision approaches.~\citet{levie2021radiounet} applies cascaded UNet modules, attempting to capture the spatial-dependent signal propagation pattern in the scene. 
To reduce data \& model complexity,~\citet{petrosyanu} proposed using an electromagnetic field (EMF)-informed loss, combined with data supervision for physics-aware learning. 
\citet{jiang2024physics} combines the aforementioned methods, reusing a similar neural architecture as~\citet{levie2021radiounet}, while designing a sophisticated, environment parameter-dependent loss. The loss is formulated based on the Volume Integral Equation (VIE), based on Maxwell’s Equations, applying supervision to the scattering portion of EMF. Such physics-informed loss is parameterized by high-precision material information within the scenario, which is inaccessible at scale, as noted by~\citet{hoydis2024learning}.

% Equivariance
Equivariance is one of the salient priors for signal propagation.
According to beamforming conditions, network geometries (antenna azimuth, tile degree, and pattern), and the transmitter-receiver distance, pathloss estimation is usually \textit{geo-coordinate-dependent} but \textit{orientation-agnostic}.
Thus, the exploration of encoding \textit{orientation-agnosticality} as a neural architecture's equivariance that facilitates radio propagation learning has been a recent trend.
One significant development is to encode the \textit{orientation-agnosticality} as wireless ray-tracing for reproducible channel estimation~\cite{Jin2505:SANDWICH, wang2024graph}, which formulates the radio propagation as \textit{shoot-n-bounce sequences} that are non-directional. 
\citet{Jin2303:Learning} models the inter-cell radio propagation interference as a relational prior that forms a graph topology in a multi-cell scenario. Besides achieving state-of-the-art performance, this method fails to provide high-fidelity pixel-level pathloss estimation. 

% Our method
Similarly to previous works, we consider that the learning kernel should be formulated under such steerable equivalence~\cite{cohen2016group}, instead of serving equivalence as part of supervision in the form of sophisticated EMF-informed loss functions~\cite{jiang2024physics,petrosyanu} or explicitly formulating equivalence in the input space~\cite{Jin2303:Learning,Jin2505:SANDWICH}. Combining recent works~\cite{levie2021radiounet} on learning (spatial) representation of radio propagation, we consider this work to be the first to report group equivariance in pathloss estimation, to the best of our knowledge.

\section{Proposed Method}
\label{sec:method}
\begin{figure*}[htbp]
  \begin{center}
    \includegraphics[width=450px]{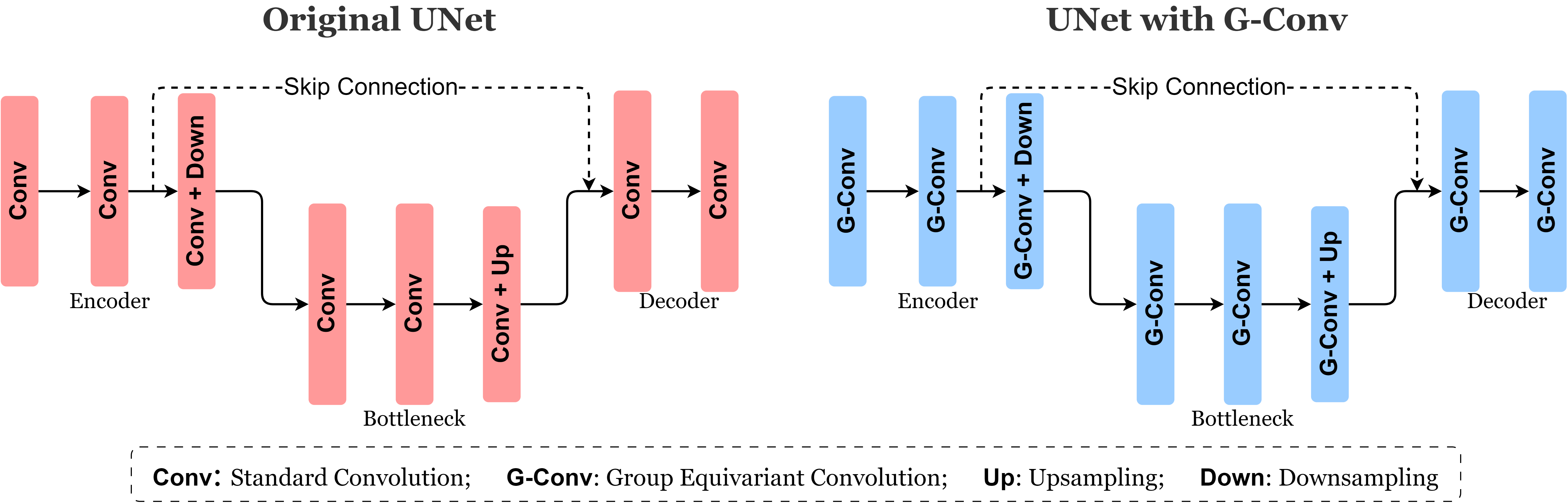}
  \end{center}
  \caption{Original UNet architecture (adapted from~\cite{ronneberger2015u} (left) and enhanced UNet with G-Conv (right)}
  \label{fig:UNet}
\end{figure*}

In this work, we propose \textbf{RadioGUNet}, a neural network architecture inspired by RadioUNet~\cite{levie2021radiounet}, a well-known UNet-based neural network for pathloss estimation. We extend underlying convolution operations with \textsc{g-conv} applied in different groups, which represent transformations such as rotation, reflection, and other symmetries. Although we focus solely on RadioUNet, the proposed extension can be applied to any convolution-based neural network for pathloss estimation, and it is not limited to UNets. The main intuition behind the use of \textsc{g-conv} is that, as signals propagate, they may rotate and reflect as they interact with different obstacles. Regular convolution is limited to translation invariance, which only captures signal propagation in free space. \textsc{g-conv}, on the other hand, can capture richer interactions with obstacles without requiring enhanced datasets.

A typical UNet (illustrated on the left side of Figure ~\ref{fig:UNet}), such as the one employed by RadioUNet, consists of an encoder/decoder type of architecture. The encoder, performs feature extraction by using small filters based on convolution, activation functions to add non-linearity, and pooling to shrink the input image while maintaining key information. The output of the encoder is also known as the bottleneck layer, which compresses/abstracts the most important information produced by the encoder and communicates it to the decoder. The decoder is the second important element of the UNet, and it uses upsampling to increase the size of the encoder's output back to its original size. Moreover, through skip connections, it uses information from the encoder that may have been discarded by the bottleneck layer. Finally, convolution is used to further refine the model's output. 

In RadioUNet, the model is trained using as input an image of the layout and the location of the transmission source in 2D coordinates. The output of the model is a propagation map where, for every pixel, coverage is estimated. In our work, we preserve the same input and output. However, we replace convolution with \textsc{g-conv}, which allows for capturing transformations of input data (i.e., rotation of the transmission source) by rotating the feature maps produced by the model, and without the need for rotating the original dataset. 

\subsection{Group equivariant convolution}
From the perspective of a UNet, the problem of pathloss estimation is that of mapping an image of a city and transmitters to that of a radio map. UNets, and particularly convolution, are \textit{translation-invariant}, which means that shifting the input image would also shift the output. This has been observed in~\cite{levie2021radiounet} and explains why convolution is an effective process for pathloss estimation. However, convolution is limited to translation invariance and does not support other transformations, such as rotations, reflections, scaling, and other symmetries which also occur as radio signals propagate from a transmitter. Group equivariant convolution, \textsc{g-conv},  generalizes convolution to larger symmetry groups beyond translation. A group is a mathematical structure capturing translations (or combinations), reflections, rotations, and scaling. 

More formally, the convolution of a filter $\psi$ for a group $G$ with a feature map $f: G\leftarrow \mathbb{R}$ is defined in ~\ref{eq:g-conv} as the sum over all elements in $G$:

\begin{equation}
\label{eq:g-conv}
[f * \psi](g) = \sum_{h \in G} f(h) \, \psi(g^{-1} h)
\end{equation}

A transformation applied to an element g on $h\in{G}$ is expressed as $gh$ while the inverse as $g^{-1}h$. Translation or shifting an input can be expressed as $ gx=x\oplus g$ and $g^{-1} x=x\ominus g$, which yields regular convolution. 

Given this formulation, more advanced groups, such as rotation, reflection, and symmetry, are shown~\cite{cohen2016group} to be equivariant, thus permitting us to apply these operations directly to the future map without the need to augment the input dataset.

\subsection{RadioGUNet}
\label{gunet}

The RadioGUNet models are constructed by adapting the base architecture in Table~\ref{tab:base_UNet} to incorporate \textsc{g-conv} operations (illustrated in the right-hand side of Figure~\ref{fig:UNet}). To study how different groups perform in pathloss estimation, different RadioGUNet variants have been created, corresponding to different symmetry groups. Specifically, the \textit{cyclic} groups ($C_2$, $C_4$, and $C_8$) and \textit{dihedral} groups ($D_2$, $D_4$, and $D_8$). The \textit{cyclic} groups ($C_n$) model rotational symmetries. The \textit{dihedral} groups ($D_n$) extend cyclic groups by incorporating reflection symmetries, thereby capturing a larger set of transformations. In both cases, the group order $n$ determines the increment of discrete rotations, i.e., rotations in steps of $360^\circ/n$. Higher-order groups, e.g., $D_8$, further increase the expressive capacity of the network at the cost of additional computational complexity and a higher number of parameters.

Unlike the RadioUNet baseline~\cite{levie2021radiounet}, which employs a cascade of two UNets, RadioGUNet utilizes a single UNet. This allows for a streamlined architecture while still being a comparable baseline. RadioGUNet is implemented using the \texttt{e2cnn}~\cite{e2cnn} framework. To ensure a fair comparison with RadioUNet, we closely follow the original training setup and evaluation protocol, including the same number of training iterations, learning rate schedules, and dataset splits. In this way, differences in results can be attributed to the model architecture rather than discrepancies in the experimental procedure.

\begin{table*}[ht!]
\centering
\caption{Summerization of Base UNet Architecture for RadioGUNet: \textit{Number of Input/Output Channels}:  indicates the dimensionality of the feature maps entering and exiting each layer. \textit{Operation} specifies the sequence of \textsc{G-Conv} layers and pooling or upsampling operations applied in each block. \textit{Activation} shows the non-linear function applied after each block.}
\begin{tabularx}{\textwidth}{|X|c|c|c|c|}
\hline
\textbf{Layer} & \textbf{Number of Input Channels} & \textbf{Number of Output Channels} & \textbf{Operation} & \textbf{Activation} \\
\hline
\multicolumn{5}{|c|}{\textbf{Encoder}} \\
\hline
Input & - & 2 (w/o Cars) / 3 (with Cars) & Input layer & - \\
\hline
Enc Block 1 & 2 (w/o Cars) / 3 (with Cars) & 6 & $2\times$\textsc{G-Conv3}$\times3$ + \textsc{MaxPool}$2\times2$ & \textsc{ReLU} \\
\hline
Enc Block 2 & $6$ & $50$ & $2\times$\textsc{G-Conv3}$\times3$ + \textsc{MaxPool}$2\times2$ & \textsc{ReLU} \\
\hline
Enc Block 3 & $50$ & $100$ & $2\times$\textsc{G-Conv3}$\times3$ + \textsc{MaxPool}$2\times2$ & \textsc{ReLU} \\
\hline
Enc Block 4 & $100$ & $100$ & $2\times$\textsc{G-Conv3}$\times3$ + \textsc{MaxPool}$2\times2$ & \textsc{ReLU} \\
\hline
Enc Block 5 & $100$ & $170$ & $2\times$\textsc{G-Conv3}$\times3$ + \textsc{MaxPool}$2\times2$ & \textsc{ReLU} \\
\hline
\multicolumn{5}{|c|}{\textbf{Bottleneck}} \\
\hline
Bottleneck & 170 & $170$ & $3\times$\textsc{G-Conv3}$\times3$ & \textsc{ReLU} \\
\hline
\multicolumn{5}{|c|}{\textbf{Decoder}} \\
\hline
Dec Block 1 & $170+170$ & $100$ & Upsample + Skip Connection + $2\times$\textsc{G-Conv3}$\times3$ & \textsc{ReLU} \\
\hline
Dec Block 2 & $100+100$ & $100$ & Upsample + Skip Connection + $2\times$\textsc{G-Conv3}$\times3$ & \textsc{ReLU} \\
\hline
Dec Block 3 & $100+100$ & $50$ & Upsample + Skip Connection + $2\times$\textsc{G-Conv3}$\times3$ & \textsc{ReLU} \\
\hline
Dec Block 4 & $50+50$ & $6$ & Upsample + Skip Connection + $2\times$\textsc{G-Conv3}$\times3$ & \textsc{ReLU} \\
\hline
Output & $6$ & $1$ & \textsc{G-Conv3}$\times3$ & - \\
\hline
\end{tabularx}
\label{tab:base_UNet}
\end{table*}

\section{Dataset}
\label{sec:dataset}
Experiments are conducted on the \textbf{RadioMapSeer} dataset~\cite {yapar2024datasetpathlosstoaradio}, which contains simulated pathloss maps under \textbf{5.9~GHz} for dense urban environments. Even though 6G networks are geared towards higher frequencies - we experiment on the same dataset to perform a comparison with previous work done with RadioUNet~\cite{levie2021radiounet}. Each sample in this dataset consists of a layout map (with buildings, air, and optionally cars), a transmitter location, and a simulated pathloss map over a $256 \times 256$ meter grid at \textbf{1-meter resolution}. The dataset covers \textbf{a dynamic range of 80~\unit{\decibel}} in pathloss, ensuring that both strong and weak signals are represented.

We follow the data split from RadioUNet~\cite{levie2021radiounet}: \textbf{500 city maps for training, 100 for validation, and 100 for testing}. Models are trained on the training set, with the best checkpoint selected via validation set performance, and final results reported on the held-out test set, which is not exposed in the training process and therefore unknown to the trained model.

Three simulation settings are used in the experiments (cf. Table~\ref{tab:model_comparison_by_params}):
\begin{itemize}
    \item \textbf{DPM-simulation, w/o Cars:} Layout map (with buildings and air) and transmitter location as input to estimation models; pathloss simulated using the Dominant Path Model (DPM). This setting captures the main line-of-sight and dominant reflection paths, resulting in sharp shadow boundaries.
    \item \textbf{IRT-simulation, w/o Cars:} Layout map (with buildings and air) and transmitter location as input; pathloss simulated using Intelligent Ray Tracing (IRT). IRT models more complex multipath effects, leading to richer and smoother shadowing patterns compared to DPM.
    \item \textbf{DPM-simulation, with Cars:} Layout map (with buildings, air, and cars), and transmitter location as input; pathloss simulated using DPM with randomly placed cars. The presence of cars introduces additional obstructions, causing further attenuation and local variations in the shadow regions.
\end{itemize}

\begin{figure}[htbp]
    \centering
    \begin{subfigure}[b]{0.31\columnwidth}
        \includegraphics[width=\linewidth]{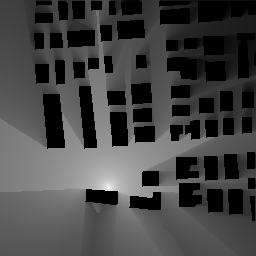}
        \caption{DPM, w/o Cars}
    \end{subfigure}
    \hfill
    \begin{subfigure}[b]{0.31\columnwidth}
        \includegraphics[width=\linewidth]{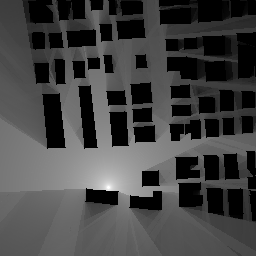}
        \caption{IRT, w/o Cars}
    \end{subfigure}
    \hfill
    \begin{subfigure}[b]{0.31\columnwidth}
        \includegraphics[width=\linewidth]{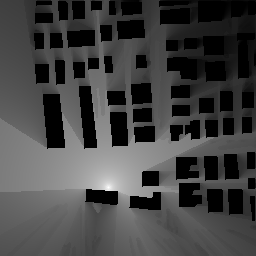}
        \caption{DPM, with Cars}
    \end{subfigure}
    \caption{Example visualized pathloss maps from the three simulation settings in RadioMapSeer. Buildings are in black, and lighter pixels indicate lower pathloss.}
    \label{fig:dataset_examples}
\end{figure}

\section{Results}
\label{sec:result}
Model performance is evaluated in terms of estimation error, quantified by the root mean squared error (RMSE)~\ref{eq:rmse} and normalized mean squared error (NMSE)~\ref{eq:nmse}, defined as follows:

\begin{equation}
\label{eq:rmse}
    \text{RMSE} = \sqrt{\frac{1}{N} \| \hat{\mathbf{PL}} - \mathbf{PL} \|_2^2}
\end{equation}

\begin{equation}
\label{eq:nmse}
    \text{NMSE} = \frac{\| \hat{\mathbf{PL}} - \mathbf{PL} \|_2^2}{\| \mathbf{PL} \|_2^2}
\end{equation}

where $\mathbf{PL}$ denotes the vector of ground-truth pathloss map, $\hat{\mathbf{PL}}$ denotes the predicted pathloss map, $\|\cdot\|_2$ is the \textsc{L2} norm, and $N$ is the total number of samples.

\begin{table*}[htbp!]
\caption{Performance Comparison of Different Models on RadioMapSeer Dataset (Ordered by Parameters)}
\begin{center}
\begin{tabularx}{\textwidth}{|l|*{6}{>{\centering\arraybackslash}X|}c|}
\hline
\multirow{3}{*}{\textbf{Model}} & \multicolumn{6}{|c|}{\textbf{Simulation Results}} & \multirow{3}{*}{\textbf{Number of Parameters (M)}} \\
\cline{2-7}
 & \multicolumn{2}{|c|}{\textbf{DPM-simulation, w/o Cars}} & \multicolumn{2}{|c|}{\textbf{IRT-simulation, w/o Cars}} & \multicolumn{2}{|c|}{\textbf{DPM-simulation, with Cars}} & \\
\cline{2-7}
 & \textbf{RMSE (\unit{\decibel})} & \textbf{NMSE} & \textbf{RMSE (\unit{\decibel})} & \textbf{NMSE} & \textbf{RMSE (\unit{\decibel})} & \textbf{NMSE} & \\
\hline
RadioGUNet-$C_2$ & $1.744$ \unit{\decibel} & $0.0090$ & $2.632$ \unit{\decibel} & $0.0230$ & $1.896$ \unit{\decibel} & $0.0115$ & 5.9 \\
\hline
RadioGUNet-$D_2$ & $1.576$ \unit{\decibel} & $0.0074$ & $2.304$ \unit{\decibel} & $0.0176$ & $1.664$~\unit{\decibel} & $0.0088$ & $11.8$ \\
\hline
RadioGUNet-$C_4$ & $1.488$ \unit{\decibel} & $0.0066$ & $2.256$ \unit{\decibel} & $0.0169$ & $1.576$ \unit{\decibel} & $0.0079$ & $11.8$ \\
\hline
RadioUNet$^{\dagger}$ & $1.600$ \unit{\decibel}$^{\dagger}$ & $0.0075$$^{\dagger}$ & $2.560$ \unit{\decibel}$^{\dagger}$ & $0.0219$$^{\dagger}$ & $1.656$ \unit{\decibel}$^{\dagger}$ & $0.0092$$^{\dagger}$ & $13.3$ \\
\hline
RadioGUNet-$D_4$ & $1.376$ \unit{\decibel} & $0.0056$ & $2.064$ \unit{\decibel} & $0.0142$ & $1.440$ \unit{\decibel} & $0.0067$ & $23.7$ \\
\hline
RadioGUNet-$C_8$ & $1.408$ \unit{\decibel} & $0.0059$ & $2.104$ \unit{\decibel} & $0.0148$ & $1.456$ \unit{\decibel} & $0.0068$ & $23.7$ \\
\hline
RadioUNet-large & $1.496$ \unit{\decibel} & $0.0066$ & $2.472$ \unit{\decibel} & $0.0203$ & $1.624$ \unit{\decibel} & $0.0084$ & $29.9$ \\
\hline
RadioGUNet-$D_8$ & \textbf{$1.304$ \unit{\decibel}} & \textbf{$0.0051$} & \textbf{$1.936$ \unit{\decibel}} & \textbf{$0.0125$} & \textbf{$1.392$ \unit{\decibel}} & \textbf{$0.0062$} & $47.3$ \\
\hline
\multicolumn{8}{l}{$^{\dagger}$Results extracted from Figure 3 of the original RadioUNet paper ~\cite{levie2021radiounet}.} \\
\end{tabularx}
\label{tab:model_comparison_by_params}
\end{center}
\end{table*}

The comparison results in Table~\ref{tab:model_comparison_by_params} highlight the effectiveness of incorporating \textsc{g-conv} for pathloss estimation compared to the baseline RadioUNet. Notably, the gains from RadioGUNet are not simply due to increased model size, as shown by comparing models with similar numbers of parameters.

For the DPM-simulation w/o cars, RadioGUNet variants with both small and moderate group orders outperform their RadioUNet counterparts. For example, RadioGUNet-$D_2$ and RadioGUNet-$C_4$ ($11.8$M parameters) achieve RMSEs of $1.576$ \unit{\decibel} and $1.488$ \unit{\decibel}, respectively, both lower than RadioUNet† ($1.600$ \unit{\decibel}, $13.3$M), and close to the much larger RadioUNet-large ($1.496$ \unit{\decibel}, $29.9$M). Higher-order models like RadioGUNet-$D_4$ ($23.7$M) further reduce the RMSE to $1.376$ \unit{\decibel}. Introducing group equivariance consistently improves estimation accuracy over standard UNets in estimating the dominant signals.

In the IRT-simulation w/o cars, the advantage of RadioGUNet becomes more pronounced. RadioGUNet-$D_2$ and $C_4$ also outperform RadioUNet† (RMSE: $2.304$/$2.256$ \unit{\decibel} vs. $2.560$ \unit{\decibel}), and RadioGUNet-$D_4$ achieves an RMSE of $2.064$ \unit{\decibel}, substantially surpassing both RadioUNet† and RadioUNet-large ($2.472$ \unit{\decibel}). This demonstrates that group equivariance helps capture intricate multipath effects, enhancing model generalization in challenging scenarios.

For the DPM-simulation with cars, representing a more dynamic environment, the improvement is consistent across model sizes. Notably, RadioGUNet-$D_2$ and $C_4$ (RMSEs $1.664$/$1.576$ \unit{\decibel}) outperform RadioUNet† ($1.656$ \unit{\decibel}), while higher-order models such as RadioGUNet-$D_4$ again provide further gains (RMSE $1.440$ \unit{\decibel}). These results emphasize the robustness of RadioGUNet variants to environmental variations.

\subsection{Discussion}

The results demonstrate that group equivariance can improve pathloss estimation across diverse propagation scenarios. RadioGUNet variants consistently achieve lower RMSE and NMSE compared to baseline models. These gains are most evident in complex conditions, such as IRT simulations or environments with vehicle obstructions, indicating that \textsc{g-conv} helps capture orientation-sensitive features and spatial correlations more effectively.

Importantly, our proposed RadioGUNet architecture utilizes a single UNet structure, in contrast to the dual-UNet cascade approach in RadioUNet. The fact that RadioGUNet achieves superior performance using a simpler architecture underscores the effectiveness of incorporating \textsc{g-conv}. These results highlight that the observed improvements in pathloss estimation are primarily due to architectural enhancements rather than increased model parameter count. This positions RadioGUNet as an efficient and highly effective solution for accurate and reliable pathloss estimation in 6G networks.

The architectural inductive bias introduced by group equivariance allows the model to generalize better to unseen layouts, suggesting strong potential for symmetry-informed learning in wireless channel modeling.

However, several limitations must be acknowledged. Training RadioGUNet models is substantially slower due to the computational overhead of current implementations, particularly those relying on the \texttt{e2cnn}~\cite{e2cnn} library. Additionally, training is occasionally unstable, with some training iterations stagnating at suboptimal performance and necessitating retraining, highlighting the need for more efficient and robust training frameworks for group equivariant models. Furthermore, RadioGUNet assumes that the underlying environment exhibits the same discrete rotational and reflection symmetries everywhere (i.e., symmetry is globally consistent). In irregular or indoor environments where such symmetries do not hold uniformly, the benefit of strict group equivariance may be reduced.

\section{Conclusions and Future Work}
\label{sec:conclusions}
% Future work could explore continuous symmetry extensions via steerable CNNs~\cite{cohen_steerable_2016}, as well as adaptive symmetry mechanisms in 3D layouts with more inconsistent mesh structures. 

Densification requirements for 6G networks call for attention to mechanisms that can improve the process of selecting where to place transmitters. To that end, pathloss estimation becomes an important problem to solve. In this paper, we show how pathloss estimation can be improved by introducing RadioGUNet, a UNet model leveraging \textsc{g-conv} to enhance pathloss estimation accuracy. Experimental results indicate that incorporating symmetry-aware operations (i.e., rotations and reflections) yields substantial gains without increasing training data size. Specifically, we demonstrate that the proposed RadioGUNet architecture consistently outperforms the baseline RadioUNet in various simulated environments, given similar numbers of model parameters.

Several promising directions exist for future research. First, the scope of this research is limited to the discrete groups. A natural extension is to extend this work to continuous symmetry groups via steerable convolutions~\cite{cohen_steerable_2016}, possibly allowing a more accurate modeling of continuously varying propagation phenomena in wireless communications. Second, given the demonstrated benefits of symmetry-aware neural networks in pathloss estimation, investigating their effectiveness in other wireless communication tasks that exhibit native equivariance, such as learning antenna pattern from wireless raytracing result~\cite{hoydis2024learning}, aligning RF fingerprinting, and interference management~\cite{Jin2303:Learning} across inter-site in the same scenario, represents an exciting avenue for further work.

\printbibliography

\end{document}